\documentstyle{article}\begin{document}
\title{Relativistic diffusion of particles with a continuous mass spectrum}
\author{ Z. Haba\\
Institute of Theoretical Physics, University of Wroclaw,\\ 50-204
Wroclaw, Plac Maxa Borna 9,
Poland\\email:zhab@ift.uni.wroc.pl}\date{\today}\maketitle
\begin{abstract}We discuss general positivity conditions
necessary for a definition of a relativistic diffusion on the
phase space. We show that Lorentz covariant random vector fields
on the forward cone $p^{2}\geq 0$ lead to a definition of a
generator of Lorentz covariant diffusions. We discuss in more
detail diffusions arising from particle dynamics in a random
electromagnetic field approximating the quantum field at finite
temperature. We develop statistical mechanics of a gas of
diffusing particles. We discuss viscosity of such a gas in an
expansion of the energy momentum tensor in gradients of the fluid
velocity.\end{abstract}
 \section{Introduction}An approximation of the dynamics of a
 non-relativistic particle in an environment of  other particles or
 random fields by a diffusion process has a long history of success \cite{pitaj}\cite{chandra}.
A direct generalization of the definition to the diffusion in
configuration Minkowski space does not work because the Minkowski
space does not have a non-negative scalar product
\cite{lopuch}\cite{hakim} (see the reviews in
\cite{deb1}\cite{hang}). One could approach the problem of finding a
realization of the diffusion by means of random dynamics
\cite{chandra}\cite{kampen} . Kramers diffusion in the phase space
can be obtained by means of a random perturbation of Hamiltonian
dynamics \cite{kampen}\cite{fish}\cite{kesten}. A relativistic
generalization of the Kramers diffusion is uniquely determined by
the requirement that the diffusion process should stay on the
mass-shell. Then, its generator must be the Laplace-Beltrami
operator on the mass-shell \cite{schay}\cite{dudley}. In this paper
we relax the mass-shell condition admitting a continuous mass
spectrum (see some earlier papers on diffusions in four-dimensional
momentum space \cite{deb3}\cite{deb2}). In fact, only stable free
particles have a definite mass in relativistic quantum field theory.
Composite particles and resonances would have a certain mass
distribution. When we give up the mass-shell condition then the
manifold of possible relativistic diffusions is increasing
substantially ( a transport theory of quantum particles with a
continuous mass spectrum has been initiated in \cite{mrow}, see also
\cite{zachar}).  Any Lorentz covariant positive definite matrix
defines a diffusion. However, if the diffusion matrix is to depend
only on the momentum $p^{\mu}$ then its arbitrariness is restricted
to a scalar function of $p^{2}$. Such diffusions have no equilibrium
( they are analogs of the Brownian motion). On general physical
grounds the equilibrium depends on the frame of reference
\cite{cmb}\cite{kujawski}\cite{matsui}. We can describe the frame by
a time-like unit vector $w^{\mu}$ . If the diffusion matrix can
depend on both $w$ and $p$ then the set of diffusions is much
larger. We determine the diffusions resulting from the requirement
that they should have the covariant (depending on $p$ and $w$)
J\"uttner equilibrium distribution \cite{juttner}.

The plan of the paper is the following. In the next section we
discuss a positivity condition resulting from the assumption that
the diffusion is defined by a probability measure and is
relativistic invariant. An unexpected relation of the diffusion
matrix to the energy momentum tensor is pointed out. This relation
is suggesting methods for a construction of relativistic diffusions.
In this section we introduce still another method of finding a
positive definite and Lorentz covariant diffusion matrix: the random
dynamics. The square of a Lorentz covariant vector field defines a
positive definite diffusion matrix. If $\Phi$ is the probability
distribution of the diffusion process then it defines a current
$N^{\mu}$ and the energy momentum tensor of a stream of diffusing
particles. In sec.3 we investigate whether these (vector and tensor)
currents can be conserved in analogy to the non-relativistic
Brownian motion (in the momentum space). We determine the diffusion
matrix and the drift of such conservative diffusions. In sec.4 we
briefly discuss a relation of the expectation value of the square of
a random vector field to the random dynamics ( Kubo's diffusion
approximation \cite{kubo1}\cite{kubo2}). In sec.5 we consider a
particle motion in a random scalar and electromagnetic fields. We
calculate an expectation value of the square of the vector field
defining the random dynamics. We show that the diffusion generator
determined by the random dynamics depends only on the field
two-point correlation function at coinciding points. The correlation
functions at finite temperature depend on the reference frame
described by the unite time-like vector $w^{\mu}$. Using the vector
$w^{\mu}$ we can define a drift by means of an expectation value of
the electromagnetic field $F_{\mu\nu}$. We have introduced such a
drift in our earlier papers \cite{habapre}\cite{habaphysica} as a
friction which brings the diffusion to the J\"uttner equilibrium on
the basis of the detailed balance condition (a generalization of the
Ornstein-Uhlenbeck process). In sec.5 we show that the diffusion
process with a continuous mass spectrum can be decomposed into the
processes on the mass-shell introduced first by Schay \cite{schay}
and Dudley\cite{dudley}. In sec.7 we introduce some notions of
(non-equilibrium) statistical physics (energy, entropy, free energy)
for a stream of diffusing particles and investigate their time
evolution and a relation between them. In secs.7 and 8 we initiate a
hydrodynamic description of a stream of relativistic diffusing
particles in analogy to the well-known description resulting from
the Boltzmann equation \cite{cerci}\cite{chapman}\cite{huang}. We
show that the diffusion equation leads to hydrodynamic equations
similar to the ones of the Boltzmann fluid. As a consequence we can
apply the standard methods of an expansion in gradients
\cite{chapman}\cite{huang} for an approximation of these equations
by a viscous (Navier-Stokes) flow.

\section{Some general positivity conditions on the diffusion tensor}

We begin with a general consideration of  requirements imposed on
the relativistic diffusion. Explicitly Lorentz invariant
 evolution  can be formulated in terms of the proper time.
 Diffusion can be considered as classical dynamics in a random field.
 In such an approach  the proper time arises in the description
 of the relativistic diffusion. Assume that there is a
transition function $P_{\tau}(x,p;x^{\prime},p^{\prime})$ in the
proper time $\tau$. Then, the diffusion generator (restricted to
the momentum variables) is
\begin{equation} \begin{array}{l}{\cal A}=d^{\mu\nu}\partial_{\mu}\partial_{\nu}+
b^{\mu}\partial_{\mu}=\partial_{\mu}(
d^{\mu\nu}\partial_{\nu}+b^{\mu})-
\partial_{\mu}b^{\mu}-(\partial_{\nu}
d^{\mu\nu})\partial_{\mu}\end{array}
\end{equation}(we denote the derivatives over $p^{\mu}$ by $\partial_{\mu}$ and
over $x^{\mu}$ by $\partial_{\mu}^{x}$). In eq.(1) we use two
alternative ways to write the generator of the diffusion which will
be discussed later on. From the definition of the diffusion matrix
\begin{equation}\begin{array}{l}
d^{\mu\nu}=\lim_{\tau\rightarrow 0}\langle
\tau^{-1}(p^{\mu}(\tau)-p^{\mu}(0))(p^{\nu}(\tau)-p^{\nu}(0))\rangle\cr
=\lim_{\tau\rightarrow 0}\tau^{-1}\int
dp^{\prime}P_{\tau}(x,p;x^{\prime},p^{\prime})
(p^{\prime\mu}-p^{\mu})(p^{\prime\nu}-p^{\nu}),\end{array}\end{equation}
where $\langle .. \rangle$ denotes a probabilistic expectation
value. It follows
\begin{equation}\begin{array}{l}a_{\mu}a_{\nu}d^{\mu\nu}=\lim_{\tau\rightarrow 0}
\langle a_{\mu}a_{\nu}
\tau^{-1}(p^{\mu}(\tau)-p^{\mu}(0))(p^{\nu}(\tau)-p^{\nu}(0))\rangle\cr
=\lim_{\tau\rightarrow 0}\tau^{-1}\langle
\Big(a(p(\tau)-p(0))\Big)^{2}\rangle\geq 0.\end{array}
\end{equation}
If the process is Lorentz invariant then the generator (1) must be
built from Lorentz tensors. It is easy to see that a homogeneous
($d^{\mu\nu}$ independent of $p$) process does not exist
 if $d^{\mu\nu}$ is to be built solely  from the Minkowski metric
$\eta^{\mu\nu}=(1,-1,-1,-1)$ because $\eta^{\mu\nu}$ is not positive
definite. Eqs.(1)-(3) hold true if $\tau=x_{0}$ (the coordinate
time) and $x\rightarrow {\bf x}$ (only the space coordinates
diffuse) but the Lorentz invariance is not explicit. A formulation
of the Lorentz invariance is more complicated in this case.

We may extend the  search for relativistic diffusions to generally
covariant theories. We note that from equations of general
relativity the Einstein tensor
\begin{equation}G_{\mu\nu}=R_{\mu\nu}-\frac{1}{2}g_{\mu\nu}R=\theta_{\mu\nu},\end{equation}
where $R_{\mu\nu}$ is the Ricci tensor and $g_{\mu\nu}$ is a
pseudoRiemannian metric, satisfies the positivity requirement
\begin{displaymath}\begin{array}{l}a_{\mu}a_{\nu}G^{\mu\nu}\geq 0,\end{array}
\end{displaymath}
if the energy momentum $\theta_{\mu\nu}$ is positive definite .
Then, both the Einstein tensor and $\theta_{\mu\nu}$ could be
chosen as the diffusion tensors. As an example of
$\theta_{\mu\nu}$ we could take the one of an ideal fluid
(discussed further in this paper) or the energy form of a map
$\phi$:$M \rightarrow {\cal M}$ between Riemannian manifolds
\cite{eells}
\begin{equation}
\theta_{\mu\nu}=g_{ab}(\phi)\partial_{\mu}\phi^{a}\partial_{\nu}\phi^{b},
\end{equation}
where $M$ is the Minkowski space-time and ${\cal M}$ is the
Riemannian manifold equipped  with the metric $g_{ab}$.

A general positive definite matrix $T^{\mu\nu}$ has the
representation \cite{gelfand}\begin{equation} T^{\mu\nu}=\int
dk\tilde{G}(k)k^{\mu}k^{\nu},
\end{equation} where $\tilde{G}(k)\geq 0$.
In the model (5) we obtain
\begin{equation}
\tilde{G}\delta(k+k^{\prime})=\langle
g_{ab}(\phi)\tilde{\phi}^{a}(k)\tilde{\phi}^{b}(k^{\prime})\rangle,
\end{equation}
if the expectation value $\langle ..\rangle$ is translation
invariant ($\tilde{\phi}$ denotes the Fourier transform of
$\phi$).

If $\tilde{G}$ depends only on $k$ and is Lorentz invariant then
the argument that a positive definite $d^{\mu\nu}$ does not exist
(as discussed after eq.(3)) still applies. We cannot define
$T^{\mu\nu}$ by means of eq.(6)using a Lorentz invariant
$\tilde{G}(k)$ function of $k$ because for such $\tilde{G}(k)$ the
integral (6) would be divergent. We shall assume that
$\tilde{G}(k,w)$ is a Lorentz invariant function of $k$ and an
additional vector $w\in V_{+}$ (without the loss of generality we
may assume $w^{2}=1$). In such a case the tensor $T^{\mu\nu}$ must
be of the form
\begin{equation}
T^{\mu\nu}(x)=\alpha(x)
w^{\mu}(x)w^{\nu}(x)-\omega(x)(\eta^{\mu\nu}-w^{\mu}(x)w^{\nu}(x)),
\end{equation}
where $\eta^{\mu\nu}$ is the Minkowski metric, $\alpha$ and
$\omega$ are some Lorentz invariant non-negative functions. We can
express them by $T$
\begin{equation}
\alpha=\int dk\tilde{G}(wk)^{2}=w^{\mu}T_{\mu\nu}w^{\nu},
\end{equation}
\begin{equation}
\omega=\frac{1}{3}\int
dk\tilde{G}((kw)^{2}-k^{2})=\frac{1}{3}(w^{\mu}T_{\mu\nu}w^{\nu}-T^{\mu}_{\mu}).
\end{equation} If the support of $\tilde{G}$ is in $V_{+}$ then
$\alpha\geq 3\omega$ as
\begin{equation}
\alpha=3\omega+\int dk\tilde{G}k^{2} .\end{equation}
 By means of the tensor $T^{\mu\nu}$ we can define some new
diffusion tensors , e.g.,
\begin{equation}\begin{array}{l}
d_{\mu\sigma}=(-\eta_{\mu\sigma}T_{\nu\rho}+
\eta_{\mu\rho}T_{\nu\sigma}-\eta_{\nu\rho}T_{\sigma\mu}+
\eta_{\nu\sigma}T_{\mu\rho})p^{\nu}p^{\rho}.
\end{array}\end{equation}If $\tilde{G}$ is non-negative and is vanishing
for $k^{2}<0$ then eq.(12) defines  an admissible diffusion tensor
(satisfying the inequality (3)), as we show in sec.5.

There is another way to find a positive diffusion matrix. We can
find a diffusion generator assuming random dynamics. Let $A(s)$ be
a random flow (the first order differential operator) depending on
the proper time $s$. We consider
\begin{equation}
Y_{\tau}=\int_{0}^{\tau}dsA(s)
\end{equation}
Then, we define
\begin{equation}\begin{array}{l}Y^{2}= \lim_{t\rightarrow
0} Y_{\tau}^{2} \tau^{-2}=\lim_{\tau\rightarrow 0}
\tau^{-2}\frac{1}{2}\int_{0}^{\tau}\int_{0}^{\tau}dsds^{\prime}
(A(s)A(s^{\prime}) +A(s^{\prime})A(s)).\end{array}\end{equation} If
$Y=\alpha^{\mu}\partial_{\mu}$ then ($Y^{+}$ denotes an adjoint of
$Y$ in $L^{2}(dp)$)
\begin{displaymath}
\langle Y^{2}\rangle=-\langle Y^{+}Y\rangle +\langle
\alpha^{\mu}\partial_{\mu}\alpha^{\nu}\rangle\partial_{\nu}.
\end{displaymath}
The first operator on the rhs is negative definite (because
$Y^{+}Y$ is non-negative). It is of the form
$-d^{\mu\nu}\partial_{\mu}\partial_{\nu}+b^{\mu}\partial_{\mu}$.
Then, it follows from the theory of partial differential operators
of the second order ( \cite{operators},sec.12) that the matrix
$d^{\mu\nu}$ satisfies  the positivity condition (3) (because the
operator $Y^{+}Y$ must be elliptic). The operator (14) in general
is degenerate, i.e., reduced to a lower dimensional (as a rule one
dimensional) second order differential operator. In general, it is
only covariant under Lorentz transformations. If the random field
is Lorentz covariant and we take a mean value of the square in
eq.(14) then we obtain a non-degenerate invariant negative
definite second order differential operator on the phase space.
This is a good candidate for a generator of the relativistic
diffusion. The randomness may come from a classical approximation
of quantum fluctuations. In the next section we discuss some
additional conditions imposed on the diffusion equation resulting
from its probabilistic interpretation. In the subsequent sections
we show that the methods (6) and (14) are related to each other.
Moreover, the diffusion generated by $Y^{2}$ of eq.(14)
approximates random dynamics.
\section{The relativistic diffusion with a continuous mass
spectrum} We define the diffusive evolution of observables
(functions of position and momentum) by the differential equation
in the proper time
\begin{equation}
\partial_{\tau}\phi=p^{\mu}\partial_{\mu}^{x}\phi+{\cal A}\phi.
\end{equation}An evolution of the probability
distribution $\Phi$ in  the proper time $\tau$ is determined by
the duality\begin{equation} \Phi_{\tau}(\phi) =\int
dpdx\sigma(p,x)\Phi(p,x)\phi_{\tau}(p,x),\end{equation}where
$\sigma$ is a Lorentz invariant non-negative function
(distribution). The dynamics of the probability distribution
$\Phi$ in the coordinate (laboratory) time is derived from the
requirement that $\Phi$ does not depend on $\tau$. This condition
gives the transport equation
\begin{equation}
p^{\mu}\partial_{\mu}^{x}\Phi={\cal
A}^{+}\Phi=d^{\mu\nu}\partial_{\mu}\partial_{\nu}\Phi+B^{\mu}\partial_{\mu}\Phi
+Q\Phi,\end{equation} where ${\cal A}^{+}$ is the adjoint of
${\cal A}$ (1) in $L^{2}(dp\sigma)$. The functions $B$ and $Q$ can
be determined from the definition of the adjoint operator and
eq.(1). For the probabilistic interpretation we need a
conservation of the current\begin{equation} N^{\mu}=\int dp\sigma
p^{\mu}\Phi.
\end{equation}
A Lorentz invariant distribution $\sigma $ in quantum theory of
resonances is of the form
\begin{equation}
\sigma(p^{2})=\int dm^{2}\delta(p^{2}-m^{2})s(m^{2})\theta(p_{0}),
\end{equation}
where $m$ has the meaning of a mass and $\theta$ denotes the
Heaviside step function. If $\sigma$ is independent of $x$ then the
requirement $\partial_{\mu}^{x}N^{\mu}=0$ leads to the equation
\begin{equation}
\partial_{\alpha}\partial_{\rho}(d^{\alpha\rho}\sigma)-\partial_{\alpha}(B^{\alpha}\sigma)+Q\sigma
=0.
\end{equation}(under the assumptions that $\sigma d$ is vanishing
at $p=0$ and $p=\infty$, so that the boundaries do not contribute
to the  integration by parts formula).

 We assume first that the
coefficients in ${\cal A}^{+}$ depend only on  $p$. From eq.(20)
we obtain $p^{\alpha}d_{\alpha \rho}=0$ (because
$\partial_{\alpha}\sigma =\sigma^{\prime}2p_{\alpha}$
 and from eq.(20) $d^{\alpha\rho}\partial_{\rho}\sigma$ must vanish
 if eq.(20) is to be satisfied for any $\sigma(p^{2})$) .
Hence, \begin{equation}
d^{\mu\nu}=P^{\mu\rho}C_{\rho\alpha}P^{\alpha\nu}.\end{equation}
where $P^{\mu\nu}$ is the projection operator
\begin{equation}P^{\mu\nu}=\eta^{\mu\nu}-(p^{2})^{-1}p^{\mu}p^{\nu}.
\end{equation} If the diffusion  depends only on $p$ then
from the Lorentz invariance
$C_{\mu\rho}=-\eta_{\mu\rho}p^{2}\gamma$. From eq.(21)
\begin{equation}
d^{\mu\nu}=-\gamma p^{2}P^{\mu\nu}\end{equation}
and
\begin{equation} b^{\mu}=\lambda p^{\mu},
\end{equation}where $\lambda$ and $\gamma\geq 0$ are functions of $p^{2}$.
 Let us note that the matrix $d^{\mu\nu}$ (23) can be derived from eq.(6) when
\begin{displaymath}
\phi^{\mu}=(x^{\mu}-p^{\mu}px)\sqrt{\gamma}
\end{displaymath}
with $g_{ab}\rightarrow \eta_{\mu\nu}$. The matrix $d$ of eq.(23)
is positive definite if and only if the spectral condition arising
from eq.(19) is satisfied.

The current $N^{\mu}$ is conserved if (from eq.(20))
\begin{equation}Q\sigma=4\lambda\sigma+2(\lambda\sigma)^{\prime}p^{2}-
6(\gamma\sigma)^{\prime}p^{2}-12\gamma\sigma.
\end{equation}We consider also the energy momentum tensor of a diffusing
particle
\begin{equation}
\theta^{\mu\nu}=\int dp\sigma \Phi p^{\mu}p^{\nu}.
\end{equation}The energy momentum is conserved if \begin{equation}
\partial_{\alpha}\partial_{\rho}(d^{\alpha\rho}p^{\nu}\sigma)-\partial_{\alpha}(p^{\nu}b^{\alpha}\sigma)+Q\sigma
p^{\nu}=0.
\end{equation}
Explicitly, with the representation (23) and (24)
\begin{displaymath}Q\sigma=5\sigma\lambda-18\gamma
\sigma+2(\lambda\sigma)^{\prime}p^{2}-6(\gamma\sigma)^{\prime}p^{2}.
\end{displaymath}
We can see that both equations (25) and (27)(for the current
conservation and the energy momentum conservation) are satisfied if
\begin{equation}
\gamma=\frac{1}{6}\lambda\equiv \kappa^{2}a(p^{2}).
\end{equation}
In such a case
\begin{equation}{\cal
A}=-\kappa^{2}a(p^{2})p^{2}P^{\mu\nu}\partial_{\mu}\partial_{\nu}
\end{equation}
(where $a(p^{2})\geq 0$ is assumed to be a regular function at
$p^{2}=0$ vanishing at infinity). The diffusion generated by the
operator (29) is the analog of the non-relativistic (Kramers)
Brownian motion which also gives a conservation of the current and
the stress tensor $(T^{0j}=N^{j},T^{lk})$(the non-relativistic limit
of eq.(17) with ${\cal A}$ of eq.(29) coincides with the Kramers
diffusion; the diffusion of Schay \cite{schay} and Dudley
\cite{dudley} has the same non-relativistic limit but does not
preserve the energy momentum).
 If the current
and energy momentum conservations are satisfied then the transport
equation reads
\begin{equation}\begin{array}{l}
p^{\mu}\partial^{x}_{\mu}\Phi={\cal
A}^{+}\Phi=-\kappa^{2}a(p^{2})p^{2}P^{\mu\nu}\partial_{\mu}\partial_{\nu}\Phi
+6\kappa^{2}a(p^{2})p^{\mu}\partial_{\mu}\Phi+
Q\Phi,\end{array}\end{equation} where
\begin{equation}
Q=12\kappa^{2}+6\kappa^{2}\sigma^{-1}(\sigma a)^{\prime}p^{2}.
\end{equation}
The diffusion (30) has an immediate generalization to an arbitrary
metric $g^{\mu\nu}$ on a pseudoRiemannian manifold $M$. For this
purpose it is sufficient to replace $\eta^{\mu\nu}$ in the
definition of $P^{\mu\nu}$ (22) by a general metric $g^{\mu\nu}$
and add the geodesic correction
$\Gamma^{\mu}_{\nu\sigma}p^{\nu}p^{\sigma}\partial_{\mu}\Phi$
(where $\Gamma $'s are the Christoffel symbols) on the rhs of
eq.(30). In such a case the energy momentum tensor (26) of the
diffusing particles can be put on the rhs of the Einstein
equations (4). If the energy momentum is not conserved then
 a varying cosmological constant is needed in order to describe the
Riemannian  geometry of a diffusing matter \cite{calo2}.

The diffusion (30) has no equilibrium. An additional friction drift
$f^{\mu}$  in eq.(30) is needed if the diffusion is to equilibrate.
The  equilibrium state depends on the frame of reference
\cite{kujawski}\cite{matsui}. We denote the velocity of the frame by
$w^{\mu}$ (where $w^{\mu}w_{\mu}=1$). Then, the drift is to depend
on $p^{\mu}$ and on $w^{\mu}$. We consider a generalization of
eq.(30) of the form
\begin{equation}\begin{array}{l}
p^{\mu}\partial^{x}_{\mu}\Phi={\cal
A}_{f}^{+}\Phi=-\kappa^{2}a(p^{2})p^{2}P^{\mu\nu}\partial_{\mu}\partial_{\nu}\Phi
+6\kappa^{2}a(p^{2})p^{\mu}\partial_{\mu}\Phi+\sigma^{-1}\partial_{\mu}(f^{\mu}\Phi)
+Q\Phi\end{array}\end{equation} We have inserted the drift $f$ in
eq.(32) in such a way that the current conservation
$\partial_{\mu}^{x}N^{\mu}=0$ is automatically  preserved. With
the drift $f^{\mu}$ the energy momentum conservation fails
\begin{equation}
\partial_{\mu}^{x}\theta^{\mu\nu}=-\int dp\Phi f^{\nu}.
\end{equation} Eq.(33) describes an exchange  of energy with an
environment (the heat bath). The equilibrium is determined by the
requirement
\begin{equation} {\cal A}_{f}^{+}\Phi_{E}=0.
\end{equation}
${\cal A}_{f}^{+}$ can be written in the second form of eq.(1)
\begin{equation}
{\cal A}_{f}^{+}=-\sigma^{-1}\partial_{\mu}(\kappa^{2}\sigma
a(p^{2})p^{2}P^{\mu\nu}\partial_{\nu} -3\kappa^{2}\sigma
a(p^{2})p^{\mu} -f^{\mu}).\end{equation}
 It follows from eq.(35) that if
\begin{equation} f^{\mu}=\kappa^{2}\sigma
ap^{2}P^{\mu\nu}\partial_{\nu}\ln\Phi_{E}-3\kappa^{2}a\sigma
p^{\mu},
\end{equation}then  eq.(34) for the equilibrium is satisfied. In
particular, for the J\"uttner distribution \cite{juttner}
\begin{equation}\Phi_{E}=\exp(-\beta wp-Kp^{2})
\end{equation}(where $K$ is an arbitrary non-negative constant and $\beta $ has the meaning of the inverse temperature)
we have\begin{equation} f^{\mu}=-\beta\kappa^{2}a\sigma
p^{2}P^{\mu\nu}w_{\nu}-3\kappa^{2}a\sigma p^{\mu}.
\end{equation}
The divergence of the energy momentum is expressed again by the
energy momentum and the conserved current (18) if $a=\sigma=1$.
Then
\begin{equation}
\partial_{\mu}^{x}\theta^{\mu\nu}
=3\kappa^{2}N^{\nu}-\beta\kappa^{2}(\theta^{\nu\mu}w_{\mu}-w^{\nu}\theta^{\mu}_{\mu}).\end{equation}
Eq.(39) remains true on the mass-shell \cite{habamod}.  We obtain
eq.(36) in a model of  the dynamics of a relativistic particle in a
random electromagnetic field in sec.5. The non-relativistic limit of
the diffusion (36) coincides with  the Ornstein-Uhlenbeck process
\begin{displaymath}
\partial_{t}\Phi=-{\bf p}\nabla_{{\bf
x}}\Phi+\partial_{j}(\partial_{j}+\beta
(p_{j}-w_{j}))\Phi,\end{displaymath} which has a current
conservation but no energy momentum conservation (because of
friction).

Instead of adding a friction $f^{\mu}$ to eq.(30) we first could
look for  a general solution of  the positivity requirement upon
the diffusion matrix (21). We do not know a general solution of
the problem. From a study of a particle motion in a random
electromagnetic field in sec.5 we obtain
\begin{equation}
C^{\mu\nu}=-a
p^{2}\eta^{\mu\nu}+b(\eta^{\mu\nu}(pw)^{2}+p^{2}w^{\mu}w^{\nu})
\end{equation} where $a$ and $b$ are non-negative constants.
The friction term leading to the J\"uttner equilibrium (37) will
also be derived in sec.5 from a particle interaction with a random
electromagnetic field.

\section{Random dynamics}
Explicitly Lorentz invariant relativistic dynamics can be expressed
in the proper time $\tau$ \cite{landau}
\begin{equation}
\frac{dx^{\mu}}{d\tau}=p^{\mu},
\end{equation}
\begin{equation}
\frac{dp^{\mu}}{d\tau}=R^{\mu}(x,p).
\end{equation}
A function $W$ of observables evolves as
\begin{equation}
\partial_{\tau}W=p^{\mu}\partial^{x}_{\mu}W+R^{\mu}\partial_{\mu}W
\equiv(X+Y)W,
\end{equation}
where
\begin{equation} X=p^{\mu}\partial^{x}_{\mu}.
\end{equation}
The current (18) ($\Phi=W$) is conserved if
\begin{equation}
\int dp W\partial_{\mu}( R^{\mu}\sigma)=0
\end{equation}
Let
\begin{equation}
Y(s)=\exp(-sX)Y\exp(sX)
=R^{\mu}(x-ps,p)(\partial_{\mu}+s\partial_{\mu}^{x}),\end{equation}
where
\begin{displaymath}
Y=R^{\mu}\partial_{\mu} \end{displaymath} Then, the solution of
eq.(43) can be expressed in the form
\begin{equation}
W_{\tau}=\exp(\tau X)W^{I}_{\tau},
\end{equation}
where
\begin{equation}
\partial_{s}W^{I}_{s}=Y(s)W^{I}_{s}.
\end{equation}

We assume that $R^{\mu}$ are random variables . In general, we have
the cumulant expansion for the expectation value
\begin{equation}\begin{array}{l}
\langle W_{t}\rangle=\exp(\tau X)\exp\Big(\int_{0}^{t}ds \langle
Y(s)\rangle  +\frac{1}{4}\int_{0}^{t}ds\int_{0}^{s}ds^{\prime}
\langle (A(s)A(s^{\prime})+A(s^{\prime})A(s))\rangle +....\Big)W.
\end{array}\end{equation}here
\begin{displaymath}
A(s)=Y(s)-\langle Y(s)\rangle \end{displaymath} If
$[Y(s),Y(s^{\prime})]=0$ and $Y$ is a linear function of Gaussian
variables then eq.(49) is exact (with no higher order terms). The
approach of Kubo \cite{kubo1}-\cite{kubo2} approximates the random
Liouville operator on the rhs of eq.(49) by an expectation value
of its square.

 The expansion of the dynamics (43) till the second
order term reads
\begin{equation}\begin{array}{l}
\langle W_{\tau}\rangle=W+\tau p^{\mu}\partial_{\mu}^{x}W  +\tau
p^{\mu}\partial_{\mu}^{x}\int_{0}^{\tau}ds\langle Y(s)\rangle W\cr+
\int_{0}^{\tau}ds\langle Y(s)\rangle
W+\frac{1}{2}(\int_{0}^{\tau}ds\langle Y(s)\rangle)^{2}W
+\frac{1}{2}\langle(\int_{0}^{\tau}ds A(s))^{2}\rangle
W+...\end{array}\end{equation} In the expansion (50) we find the
$\tau^{2}$ term of eq.(14). There is also the first order
differential operator (of the first order in $\tau$) which can be
defined by
\begin{displaymath}
K_{1}=\lim_{\tau\rightarrow 0}\int_{0}^{\tau}ds\langle Y(s)\rangle
 \tau^{-1}.
\end{displaymath} Then, there appears in eq.(50) a term which is of the first order in derivatives
and the second order in time. It  can be defined by means of the
formula
\begin{equation}
K_{2}=\lim_{\tau\rightarrow 0}\Big(\int_{0}^{\tau}ds\langle
Y(s)\rangle-K_{1}\tau \Big)\tau^{-2} .\end{equation} $\langle
Y^{2}\rangle+K_{2}$ determines the diffusion generator (where
$Y^{2}$ is defined in eq.(14)). The expansion (49) shows that the
diffusion generated by $\langle Y^{2}\rangle+K_{2}$ is related to
random dynamics. In fact, Kubo shows \cite{kubo1}that the
$\tau^{2}$ behaviour in random dynamics at times short in
comparison to the correlation time goes into the diffusive $\tau$
behaviour at times large in comparison to the correlation time.
His argument is equivalent to the rigorous Markov approximation of
ref.\cite{kesten} which is using a time rescaling from a
microscopic time to the macroscopic time. In the next section we
calculate the expectation values in eq.(50) for a particle in a
random electromagnetic field.

\section{Motion in a random scalar and electromagnetic fields}
The simplest example of $R^{\mu}$ in eq.(42) is defined by a scalar
field $\phi$
\begin{displaymath}
R_{\mu}=\partial^{x}_{\mu}\phi.
\end{displaymath}
Then, $Y^{2}$ of eq.(14) has the form
\begin{equation}
{\cal A}=T^{\mu\nu}\partial_{\mu}\partial_{\nu},
\end{equation}
where $T^{\mu\nu}$ has been calculated in eqs.(6)-(8). We do not
develop this diffusion any further because in the rest frame ($
w=(1,0,0,0)$ and $\alpha=0$) it coincides with the Brownian motion.
 We can consider $R^{\mu}$ constructed from higher rank tensor fields and
 $p^{\mu}$. In this paper we restrict ourselves to
 \begin{equation}
 R^{\mu}={\cal R}^{\mu\nu}p_{\nu},
 \end{equation} where ${\cal R}_{\mu\nu}$ is the antisymmetric tensor
 of an electromagnetic field. The
 electromagnetic current is \begin{equation} {\cal J}_{\nu}=\partial^{\mu}{\cal
R}_{\mu\nu} \end{equation} We assume that
\begin{equation}
\langle {\cal J}^{\mu}\rangle= r w^{\mu}
\end{equation}
with a certain constant $r$. We split

 \begin{equation}
 {\cal R}_{\mu\nu}=F^{av}_{\mu\nu}+F_{\mu\nu}
 \end{equation}
 where $F^{av}$ is the mean value of ${\cal R}$. $F_{\mu\nu}$
 has zero mean value and is Poincare invariant. $rw_{\nu}$ has the meaning
 of a constant current. The current (54) results from a charge $r$
 moving with the velocity
$w^{\mu}$. We can obtain such a current  in a finite temperature
quantum field theory of interacting electromagnetic and complex
scalar fields. The density matrix is
\begin{equation}
\exp(-\beta P^{\nu}w_{\nu})
\end{equation} where $P_{\mu}$ is the four-momentum of the quantum
fields and $w^{\mu}$ describes the moving frame. Then, calculating
the expectation value (of the current (54) of the  quantum scalar
complex field $\phi$ interacting with a quantum electromagnetic
field) we obtain
\begin{equation}\begin{array}{l}\langle {\cal J}_{\mu}\rangle_{\beta}= iTr\Big(\exp(-\beta
P^{\nu}w_{\nu})(\overline{\phi}\partial_{\mu}^{x}\phi-\phi\partial_{\mu}^{x}\overline{\phi})\Big)
 \cr=\int d{\bf k}k_{0}^{-1}k_{\mu}(\exp(\beta kw)-1)^{-1}=r w_{\mu}
 \end{array}\end{equation} (with a certain constant $r$).
There is  another way to see that an introduction of a current
leads to a non-zero expectation value of $
\partial_{\sigma} {\cal R}_{\mu\nu}$ . If we add the term
$\int A_{0}d{\bf x}$ to the Hamiltonian of the quantum
electromagnetic field or in a covariant and gauge invariant way
the term $\int A_{\mu}w^{\mu}dx$ to the action (here $A_{\mu}$ is
the electromagnetic vector potential) then $\langle { \cal
R}_{\mu\nu,\sigma}\rangle\neq 0$.

 We
assume that an average $\langle\cdot \rangle$ over $F$ is defined
which preserves the Lorentz symmetry. This means that the two-point
function defined by
\begin{equation}\langle
F_{\mu\nu}(x)F_{\sigma\rho}(x^{\prime})\rangle=G_{\mu\nu
;\sigma\rho}(x-x^{\prime})
\end{equation}
is a tensor. $G_{\mu\nu ;\sigma\rho}$ is symmetric under the
exchange of indices $(\mu\nu;x)$ and $(\sigma\rho;x^{\prime})$ and
antisymmetric under the exchange $\mu\rightarrow\nu$ and
$\sigma\rightarrow\rho$. We impose the Bianchi identities
\begin{equation}{ \cal R}_{\mu\nu,\sigma}+{ \cal
R}_{\sigma\mu,\nu}+{ \cal R}_{\nu\sigma,\mu}=0\end{equation} on
the tensor field ${\cal R}$. In terms of the two-point function
\begin{equation}
\partial_{\alpha}\epsilon^{\alpha\beta\mu\nu}G_{\mu\nu
;\sigma\rho}=0.
\end{equation}
In Fourier transforms eq.(61) reads
\begin{equation}\langle\overline{\tilde{F}}_{\mu\nu}(k)\tilde{F}_{\sigma\rho}(k^{\prime})\rangle=
\tilde{G}_{\mu\nu ;\sigma\rho}(k)\delta(k-k^{\prime}),
\end{equation}
where $\tilde{G}_{\mu\nu ;\sigma\rho}(k)$ is a tensor which must
be constructed from the vectors $k_{\mu}$,$w_{\mu}$ and the
fundamental four-dimensional tensors $\eta_{\mu\rho}$ and
$\epsilon_{\mu\nu\rho\sigma}$.  Hence,in general we could have
\begin{equation}\begin{array}{l}
 \tilde{G}_{\mu\nu
 ;\sigma\rho}(k)=a_{1}(\eta_{\mu\sigma}k_{\nu}k_{\rho}-
 \eta_{\mu\rho}k_{\nu}k_{\sigma}+\eta_{\nu\rho}k_{\mu}k_{\sigma}-\eta_{\nu\sigma}k_{\mu}k_{\rho})
+a_{0}\epsilon_{\mu\nu\sigma\rho}\cr
+a_{2}(\eta_{\mu\sigma}w_{\nu}k_{\rho}-
 \eta_{\mu\rho}w_{\nu}k_{\sigma}+\eta_{\nu\rho}w_{\mu}k_{\sigma}-\eta_{\nu\sigma}w_{\mu}k_{\rho})\cr
+a_{3}(\eta_{\mu\sigma}w_{\nu}w_{\rho}-
 \eta_{\mu\rho}w_{\nu}w_{\sigma}+\eta_{\nu\rho}w_{\mu}w_{\sigma}-\eta_{\nu\sigma}w_{\mu}w_{\rho})
\end{array}\end{equation}However, the Bianchi identities (61) and the requirement of positivity
of the probability measure in eq.(59) (see \cite{habajpa}) lead to
$a_{0}=a_{2}=a_{3}=0$ (in the case of a quantum free
electromagnetic field at finite temperature we obtain also the
representation (63) where only $a_{1}\neq 0$). We have
\begin{equation}\begin{array}{l}
 G_{\mu\nu
;\sigma\rho}(x,x^{\prime})=\int dk\tilde{G}_{\mu\nu
 ;\sigma\rho}(k)\exp(ik(x-x^{\prime}))\cr
 =\int dk \tilde{G}(k)\exp (ik(x-x^{\prime}))(\eta_{\mu\sigma}k_{\nu}k_{\rho}-
 \eta_{\mu\rho}k_{\nu}k_{\sigma}+\eta_{\nu\rho}k_{\mu}k_{\sigma}-\eta_{\nu\sigma}k_{\mu}k_{\rho})
\end{array}\end{equation}
and\begin{equation} \langle
F_{\mu\nu}(x)F_{\sigma\rho}(x^{\prime})\rangle_{\beta}=-D_{\mu\nu
;\sigma\rho}G(x-x^{\prime}),\end{equation}
 where\begin{equation}\begin{array}{l}
 D_{\mu\nu;\sigma\rho}
=-\eta_{\mu\sigma}\partial_{\nu}\partial_{\rho}+
\eta_{\mu\rho}\partial_{\nu}\partial_{\sigma}-
\eta_{\nu\rho}\partial_{\sigma}\partial_{\mu}+
\eta_{\nu\sigma}\partial_{\mu}\partial_{\rho}
\end{array}\end{equation}
   The two-point
function is positive definite if and only if $\tilde{G}(k)$ in
eq.(64) satisfies the condition

 \begin{equation} \tilde{G}(k)\geq 0
\end{equation} and $\tilde{G}(k)$=0 if $k^{2}< 0$\cite{habajpa}.

 It follows from eq.(64) that
\begin{equation}\begin{array}{l} (D_{\mu\nu;\sigma\rho}G)(0)
=\eta_{\mu\sigma}T_{\nu\rho}- \eta_{\mu\rho}T_{\nu\sigma}+
\eta_{\nu\rho}T_{\sigma\mu}- \eta_{\nu\sigma}T_{\mu\rho},
\end{array}\end{equation}
where $T_{\mu\nu}$ is defined in eq.(6).

It is instructive to see the relation between $T$ and the
energy-momentum tensor $\theta_{\mu\nu}$ for the electromagnetic
field. We have
\begin{equation}
\theta_{\nu\rho}=\eta^{\mu\sigma}F_{\mu\nu}F_{\sigma\rho}-\frac{1}{4}\eta_{\nu\rho}
F_{\alpha\sigma}F^{\alpha\sigma}.
\end{equation}
 Hence, the
expectation value of the energy momentum tensor is expressed as
\begin{equation}
\langle\theta_{\mu\nu}\rangle=2T_{\mu\nu}-\frac{1}{2}\eta_{\mu\nu}T^{\rho}_{\rho}
\end{equation}

 We can show that in eq.(51)
\begin{equation}
 K_{1}=\langle {\cal R}_{\mu\nu}\rangle p^{\nu}\partial^{\mu}
 \end{equation}
and that the rhs of eq.(52) can be expressed
as\begin{equation}\begin{array}{l} \Big(\int_{0}^{\tau}ds\langle
Y(s)\rangle-K_{1}\tau \Big)\tau^{-2}
=\tau^{-2}\int_{0}^{\tau}ds\int_{0}^{s}ds^{\prime}
\partial_{\sigma}F^{av}_{\mu\nu}(x-(s-s^{\prime})p)p^{\sigma}p^{\nu}\partial_{\mu}
\end{array}\end{equation}
Then, taking the limit $\tau\rightarrow 0$ in eq.(72) we
obtain\begin{equation} K_{2}=\langle \partial_{\sigma}{\cal
R}_{\mu\nu}\rangle p^{\sigma}p^{\nu}\partial^{\mu}
 \end{equation}
We apply the covariance (65) to calculate the expectation value of
the square of the Liouville operator. In this way we calculate the
 $\tau^{2}$ term in the expansion (49). Then, ($A(s)$ is defined
below eq.(49))
\begin{equation}\begin{array}{l}Y^{2}=\lim_{\tau\rightarrow
0}\frac{1}{\tau^{2}}\langle(\int_{0}^{\tau}A(s)ds)^{2}\rangle=
\lim_{\tau\rightarrow
0}\frac{1}{2\tau^{2}}\int_{0}^{\tau}ds\int_{0}^{\tau}ds^{\prime}\langle
A(s)A(s^{\prime})+A(s^{\prime})A(s)\rangle
\cr=(\eta^{\mu\sigma}T^{\nu\rho}-\eta^{\mu\rho}T^{\nu\sigma}+\eta^{\nu\rho}T^{\mu\sigma}-\eta^{\nu\sigma}T^{\mu\rho})
p_{\nu}\partial_{\mu}p_{\rho}\partial_{\sigma}
\end{array}\end{equation}
If we apply the formulas (8)  then we obtain
\begin{equation}\begin{array}{l}
Y^{2}
=-2\omega\partial_{\mu}P^{\mu\nu}p^{2}\partial_{\nu}\cr
+(\rho+\omega)((wp)^{2}\partial_{\mu}\partial^{\mu}-2wpw^{\mu}p^{\rho}\partial_{\mu}\partial_{\rho}
+p^{2}w^{\mu}w^{\rho}\partial_{\mu}\partial_{\rho}-w^{2}p^{\nu}\partial_{\nu}
-2wpw^{\nu}\partial_{\nu})\equiv {\cal A}_{w}
\end{array}\end{equation}
The limit  in eq.(74) depends only on $DG(0)$ of eqs.(65)and (68).
It does not depend on $p_{\mu}$. The dependence on $p_{\mu}$ of
$D_{\mu\nu;\rho\sigma}G((s-s^{\prime})p)$ in eq.(49) is of
 higher than the second order in $\tau$(because after integration in eq.(49) over $s$ and $s^{\prime}$
 we obtain a factor $\tau^{3}$).

The transport equation determined by the diffusion $Y^{2}+K_{2}$
(eqs.(75) and (73)) reads (it is an extension of eqs.(21),(32) and
(40))
\begin{equation}\begin{array}{l}
p^{\mu}\partial^{x}_{\mu}\Phi={\cal
A}_{f}^{+}\Phi=\partial_{\mu}(D^{\mu}+f^{\mu})\Phi\cr
\equiv\partial_{\mu}\Big(-aP^{\mu\nu}p^{2}\partial_{\nu} -r
p^{2}P^{\mu\nu} w_{\nu}
+b((pw)^{2}\partial^{\mu}-(wp)(w^{\mu}p^{\rho}+w^{\rho}p^{\mu})
\partial_{\rho}+p^{2}w^{\mu}w^{\rho}\partial_{\rho})\Big)\Phi
\end{array}\end{equation}
Here, $a=2\omega$, $b=\rho+\omega$ and the drift (as discussed in
sec.3) resulting from eq.(73) is\begin{equation} f^{\mu}=-r
p^{2}P^{\mu\nu} w_{\nu}.
\end{equation} $r$ is an arbitrary real parameter which could be determined from the
original model of random dynamics(as in eq.(58)).

\section{Restriction to the mass-shell} First, we discuss the diffusion
at zero temperature corresponding to $w=0$ in eq. (76). The
calculations of $\langle Y^{2}\rangle$
 in sec.5 lead to the following analog of the Brownian motion
\begin{equation}\begin{array}{l}
\partial_{\tau}\Phi\equiv {\cal G}\Phi=
p^{\mu}\partial_{\mu}^{x}\Phi+ \kappa^{2}\triangle_{P}\Phi
\end{array}\end{equation} where $\kappa^{2}$ is a diffusion constant
and
\begin{equation}\triangle_{P}=-\partial_{\mu}P^{\mu\nu}p^{2}\partial_{\nu}
\end{equation} and ${\cal G}$ is an operator in the phase space  of
points $(x,p)$ such that $p^{2}\geq 0$ and $p_{0}\geq 0$ . The
operator $-\triangle_{P}$ can be defined as a symmetric positive
definite operator in $L^{2}(dp\sigma)$ (it is symmetric also in
$L^{2}(dp))$
\begin{equation}
(f_{1},\triangle_{P}f_{2})=D(P\partial f_{1},p^{2}P\partial
f_{2})=\int
dp\sigma(p)p^{2}P^{\mu\nu}\partial_{\mu}f_{1}\partial_{\nu}f_{2}.
\end{equation}

Let us consider the direct integral of Hilbert spaces
\begin{displaymath}
{\cal H}=\int dm^{2}s(m^{2}) {\cal H}_{m}
\end{displaymath}
where \begin{equation} (f_{1},f_{2})_{m}=\int d^{4}p
\delta(p^{2}-m^{2})f_{1}(p)f_{2}(p)
\end{equation}  Define the bilinear form
\begin{equation}\begin{array}{l} g(\partial f_{1},\partial f_{2})_{m}=\int d^{4}p
\delta(p^{2}-m^{2})\gamma^{jk}\partial_{j}f_{1}\partial_{k}f_{2}
\equiv (f_{1},-\triangle_{H}^{m}f_{2})_{m}\end{array}
\end{equation} with
\begin{equation} \gamma^{jk}=m^{2}\delta^{jk}+p^{j}p^{k}
\end{equation}and
\begin{equation}
\triangle_{H}^{m}=\gamma^{-\frac{1}{2}}\partial_{j}\gamma^{jk}\gamma^{\frac{1}{2}}\partial_{k}.
\end{equation}
The operator $\triangle_{H}^{m}$ is the Laplace-Beltrami operator
on the hyperboloid $p^{\mu}p_{\mu}=m^{2}$.
$\gamma=\det(\gamma_{jk})$ and $\gamma^{-1}=m^{4}p_{0}^{2}$,
$p_{0}=\sqrt{m^{2}+{\bf p}^{2}}$, $k=1,2,3$ .  The operator
$\triangle_{H}$ in Eq.(84) can be expressed in coordinates ${\bf
p}$ as
\begin{equation}\begin{array}{l} \triangle_{H}^{m}
=(\delta^{jl}m^{2}+p^{j}p^{l})\frac{\partial}{\partial
p^{l}}\frac{\partial}{\partial
p^{j}}+3p^{l}\frac{\partial}{\partial p^{l}}\end{array}
\end{equation} We have
\begin{equation} \int dm^{2}s(m^{2})g(\partial f_{1},\partial
f_{2})_{m}=-D(P\partial f_{1},p^{2}P\partial f_{2}).\end{equation}
On the lhs of  eq.(86) in each ${\cal H}_{m}$ we can express the
derivatives either by $p_{\mu}$ or by $p_{j}$ treating $p_{0}$ as
a function of $p_{j}$. The relation between the derivatives is
\begin{equation}
\tilde{\partial}_{j}=(\partial_{j}p_{0})\partial_{0}+\partial_{j}.
\end{equation}

The functions $f_{j}$ depend on a four-vector $p$. On the rhs of
eq.(86) we have an explicitly Lorentz invariant formula expressed
by Lorentz vectors and a measure which is a scalar. On the lhs the
Lorentz invariance is not explicit. The lhs is explicitly positive
definite whereas the positive definiteness of the rhs is not
obvious.

$\triangle_{H}^{m}$ is the generator of the diffusion defined
first by Schay\cite{schay} and Dudley\cite{dudley}. The drift (77)
can also be restricted to the mass-shell
\cite{habapre}-\cite{habaphysica} (it determines the exponential
speed of the decay to the equilibrium, see \cite{calo}).

\section{Thermodynamics of diffusing particles}
We are interested in this section in a thermodynamic description
of a system of diffusing particles in  models which have an
equilibrium solution. We find the equilibrium solution in the
model (76) from the requirement
\begin{equation}
(D^{\mu}-r p^{2}P^{\mu\nu}w_{\nu})\Phi_{E}=0.
 \end{equation} We obtain  the
J\"uttner equilibrium (37) if
\begin{equation}
 r=\beta (a-b)=\beta(\omega-\alpha).
 \end{equation} $\Phi_{E}$ is also an
equilibrium solution of the model (35)-(36)
  with $r=\kappa^{2}\beta$.

 From eq.(76) we obtain
\begin{equation}
\begin{array}{l}
\partial_{\mu}^{x}\Theta^{\mu\nu}=-(b-3a)N^{\nu}+r\Theta^{\mu}_{\mu}w^{\nu}-r\Theta^{\mu\nu}w_{\mu}
-2bw^{\nu}w_{\mu}N^{\mu}.\end{array}\end{equation} In the model
(35)-(36) we have $b=a=0$ on the rhs of eq.(90).

 We can introduce thermodynamic notions useful in a
description of a stream of relativistic diffusing particles. We
define the relative entropy current (Kulback-Leibler entropy; we
follow definitions of our earlier paper \cite{habamod} concerning
the model on the mass-shell)
\begin{equation}\begin{array}{l} S^{\mu}_{K}= N^{-1}\int
dp\sigma(p)p^{\mu}\Phi\ln(N^{-1}\Phi N_{E}\Phi_{E}^{-1}).
\end{array}\end{equation}Here,

\begin{equation} N=\int d{\bf x}N_{0}
\end{equation}is the charge normalization constant
($N_{E}$ in the case of the equilibrium).  It follows that
$\Omega=N^{-1}\sigma^{-1}p_{0}^{-1}\Phi$ has the meaning of the
probability density (on the mass-shell $N^{-1}\Phi$ is the
probability density). It can be shown that \begin{equation} \int
d{\bf x}S_{K}^{0}\geq 0\end{equation} and
\begin{equation}
\partial_{\mu}S^{\mu}_{K}=-N^{-1}\int
dp\sigma(p)\Phi\alpha^{\mu\nu}\partial_{\mu}\ln R\partial_{\nu}\ln
R,
\end{equation}
where $R=\Phi\Phi_{E}^{-1}$. $\alpha^{\mu\nu}$ comes from the
formula (76) where
\begin{equation} D^{\mu}=\alpha^{\mu\nu}\partial_{\nu}.
\end{equation} From positive definiteness of $-{\cal A}_{w}$ the momentum dependent coefficients
$\alpha^{\mu\nu}$ satisfy the positivity condition
$a_{\mu}a_{\nu}\alpha^{\mu\nu}\geq 0$ (see our discussion after
eq.(14)).
 It follows that $\int d{\bf x}S_{K}^{0}$ is a positive function
decreasing monotonically to zero (it can be interpreted as the
entropy of the system plus its heat bath). The entropy current of
the particle system is
\begin{equation}\begin{array}{l} S^{\mu}=- N^{-1}\int
dp\sigma(p)p^{\mu}\Phi\ln\Phi.
\end{array}\end{equation}
We have
\begin{equation}
\partial_{\mu}S^{\mu}=N^{-1}\int
dp\sigma(p)\Phi\alpha^{\mu\nu}\partial_{\mu}\ln
\Phi\partial_{\nu}\ln \Phi-3r w_{\mu}N^{\mu}.
\end{equation}   The entropy
is defined as \begin{equation}S= \int d{\bf x}S^{0}.\end{equation}
Then
\begin{equation}
\partial_{0}S=N^{-1}\int
dp\sigma(p)\Phi\alpha^{\mu\nu}\partial_{\mu}\ln
\Phi\partial_{\nu}\ln \Phi-3r N^{-1}w_{\mu}\int d{\bf x}N^{\mu}.
\end{equation}
We may choose the frame $w=(1,{\bf 0})$. Then,
\begin{equation}w^{\mu}\int d{\bf x}N_{\mu}=\int d{\bf x}N_{0}=N\geq 0
\end{equation}
So, the first term on the rhs of eq.(99) is positive whereas the
second is negative. We define the free energy
\begin{equation}
{\cal F}=\beta^{-1}N\int d{\bf
x}S_{K}^{0}-N\beta^{-1}\ln(NN_{E}^{-1})
\end{equation}
and the energy
\begin{equation}
{\cal W}=\int d{\bf x}\theta_{00}.
\end{equation}
 Then, the basic thermodynamic relation
(at fixed temperature) \begin{equation} \beta^{-1}S={\cal W}-{\cal
F}
\end{equation}
comes out  as an identity. The time evolution of each term in
eq.(103) is determined by eqs.(39),(91) and (99). From eq.(103) we
can see that the decrease of $\partial_{0}S$ in eq.(99) is caused
by the decrease of ${\cal W}$ in eq.(103).

\section{Fluid velocity and the macroscopic energy momentum}

On the basis of the diffusion theory we can develop a hydrodynamic
description of a gas of diffusing particles. We define the density
(a scalar)
\begin{equation}
\rho(\Phi)=\int dp \sigma(p)\Phi
\end{equation}
and the mean momentum\begin{equation}
v^{\mu}=\rho(\Phi)^{-1}N^{\mu}\end{equation}We can  write the
energy-momentum tensor in the form

\begin{equation} \theta^{\mu\nu}=\rho
v^{\mu}v^{\nu}+\tau^{\mu\nu}
\end{equation}
where
\begin{equation}
\tau^{\mu\nu}=\int
dp\sigma(p)(p^{\mu}-v^{\mu})(p^{\nu}-v^{\nu})\Phi\end{equation}
 Let
\begin{equation}
n^{2}=v_{\alpha}v^{\alpha}
\end{equation}
\begin{equation}
v^{\mu}=nu^{\mu}
\end{equation} Then, $u_{\mu}u^{\mu}=1$.
We define the projection operator
\begin{equation}
h^{\mu\nu}(u)=\eta^{\mu\nu}-u^{\mu}u^{\nu}
\end{equation}
such that $u_{\mu}h^{\mu\nu}=0$. In general, we can write
\begin{equation}
\tau^{\mu\nu}=-\Pi h^{\mu\nu}+\sigma^{\mu\nu}
\end{equation}(for a gas
of free relativistic particles in an equilibrium we have
$\sigma^{\mu\nu}=0$). So, we write the energy momentum tensor in
the form
\begin{equation}
\theta^{\mu\nu}==Eu^{\mu}u^{\nu}-\Pi
h^{\mu\nu}+\sigma^{\mu\nu}\end{equation}where
\begin{equation}
E=\rho n^{2} \end{equation}We shall discuss in more detail the
models (35)-(36) and (76) with $b=0$ (in these models fluid
equations have a simple interpretation). As discussed in
\cite{eckart}\cite{landaufluid}\cite{rom} the relativistic fluid
equations are projections of the conservation laws. Let us identify
the divergence equations (39)  with the equations resulting from the
definition of the energy-momentum tensor (112) (calculating
$h_{\alpha\nu}\partial_{\mu}\theta^{\mu\nu}$ in two
ways)\begin{equation}\begin{array}{l} -(E+\Pi)
u^{\mu}\partial_{\mu}u_{\alpha}+h_{\alpha\nu}\partial^{\nu}\Pi-
h_{\alpha\nu}\partial_{\mu}\sigma^{\mu\nu} \cr=\beta\kappa^{2}
h_{\alpha\nu}\Big(-\Pi w^{\nu}+w_{\mu}\sigma^{\mu\nu}-
w^{\nu}(E-3\Pi+\sigma^{\mu}_{\mu})\Big).
\end{array}\end{equation}
Let us note that the fluid equations do not depend on the
$N^{\nu}$-term in eq.(90). When there is no friction ($\beta=0$ )
we obtain the same equations as the ones of Landau and Lifshitz
\cite{landaufluid}. The non-relativistic limit of eq.(114) gives
(Euler or  Navier-Stokes depending on $\sigma^{\mu\nu}$) fluid
equations.

\section{Viscosity of relativistic diffusing particles} The velocity $w^{\mu}$
 in the transport equation (76) is an arbitrary auxiliary variable. In this section we allow
 $w^{\mu}$ to depend on $x$. We can
interpret  an equation with $x$- dependent $w$ either as the
diffusion with friction  in a  frame moving with the local velocity
$w^{\mu}(x)$ or as the diffusion of a stream of particles in a fluid
moving with the velocity $w^{\mu}(x)$ and playing the role of the
heat bath.
 In the state
\begin{equation}
\Phi_{E}=\exp(\chi(x)-\beta(x) pw(x))
\end{equation}
  the particle mean velocity
is $u^{\mu}_{E}(x)=w^{\mu}(x)$. $\Phi_{E}$ satisfies the equation
${\cal A}^{+}_{w}\Phi_{E}=0$ but the lhs of eq.(76) is different
from zero depending on the derivatives of the velocity $w$,
temperature $\beta^{-1}$ and the density $\exp(\chi(x))$. Then,
$\Phi_{E}({\bf x},x_{0})$ evolves from an initial time $x_{0}$
according to the diffusion equation (76). The initial $\Phi_{E}$ and
its subsequent time evolution constitute a useful description of a
state close to the local equilibrium. In this section we discuss the
time evolution of $\Phi_{E}$ (115) and the fluid velocity in more
detail ((115) is an analog of the local equilibrium of the
non-relativistic Boltzmann equation discussed in an elementary way
in \cite{huang} and in a more advanced form in \cite{chapman}; for
the relativistic case see \cite{is}\cite{groot} and \cite{kremer}) .

 The
mean momentum $v^{\mu}$ of the stream in the state $\Phi_{E}$ is
\begin{equation}\begin{array}{l} \rho_{E}v_{E}^{\mu}=
\int dp\sigma(p)p^{\mu}\exp(-\beta
 p_{\nu}w^{\nu}(x))= n_{E}
 \rho_{E}w^{\mu}(x),\end{array}\end{equation}where
 \begin{displaymath}\rho_{E}=\int dp\sigma(p)\exp(-\beta
 p_{\nu}w^{\nu}(x)).\end{displaymath}
$n_{E}$ is the normalization of $v^{\mu}_{E}=n_{E}w^{\mu}$. We
have\begin{equation}\begin{array}{l} n_{E} \rho_{E}=\int
dp\sigma(p)pw\exp(-\beta
 pw)=\int dm^{2}s(m^{2})n_{m}\rho_{E}^{m}\end{array}\end{equation}
where $n_{m}$ in\begin{equation}n_{m}\rho_{E}^{m}= \int
dp\delta(p^{2}-m^{2})pw\exp(-\beta
 p_{\nu}w^{\nu}(x))\end{equation}
 is the value of $n$
calculated on the mass-shell (see e.g. \cite{groot}).

In this section we consider the models (35)-(36) or  (76) with
$b=0$ discussed in sec.7. We express the transport equation as the
diffusion equation in an invariant time variable
\begin{equation}
\tau=x_{\mu}w^{\mu}
\end{equation}( different from $\tau$ in eqs.(15) and (41);
the deterministic proper time  satisfies $\tau=x_{0}$ in the
particle rest frame whereas  for the time (119) we obtain
$\tau=x_{0}$ in the observer's rest frame). We may change the
coordinates $(x_{0},{\bf x})\rightarrow (\tau,{\bf x})$ and
$(p_{0},{\bf p})\rightarrow (q,{\bf p})$ where
\begin{equation}
q=w^{\mu}p_{\mu} \end{equation} In such a case
\begin{equation}\begin{array}{l}
p_{\mu}\partial_{x}^{\mu}=q\partial_{\tau}+{\bf p}\nabla_{{\bf x}}
+p^{\mu}x_{\nu}\partial_{\mu}^{x}w^{\nu}\partial_{\tau}\equiv
q\partial_{\tau} +{\cal D}\end{array}
\end{equation}
Let us consider \begin{equation}
\Phi=\Gamma\Phi_{E}\end{equation}Then,
\begin{equation} p^{\mu}\partial^{x}_{\mu}\Gamma={\cal A}\Gamma-
(p^{\mu}\partial_{\mu}^{x}\ln\Phi_{E})\Gamma
\end{equation}
In new coordinates\begin{equation}\begin{array}{l}
q\partial_{\tau}\Gamma={\cal A}\Gamma-{\cal D}\Gamma-
(p^{\mu}\partial_{\mu}^{x}\ln\Phi_{E}) \Gamma
\end{array}\end{equation} The solution of eq.(124) can be expressed
by an exponential of $q^{-1}({\cal A}-{\cal D})$. If in the lowest
order we neglect the second order space-time derivatives and the
squares of the  first order derivatives of $w$ in $\Gamma$ then
the solution of eq.(124) reads
\begin{equation}
\Gamma(x)=1-q^{-1}\int_{0}^{\tau}ds\exp((\tau-s)q^{-1}{\cal
A})p^{\mu}\partial_{\mu}^{x}\ln\Phi_{E}(s,{\bf x})
\end{equation}
Taking only the lowest  eigenvalue $\nu>0$ of $-q^{-1}{\cal A}$ in
eq.(125) we obtain the relativistic relaxation time approximation
from the formula
\begin{equation}
\int_{0}^{\tau} ds\exp(-\nu(\tau-s)) F(s)\simeq \nu^{-1}F(\tau)
\end{equation}
true for large $\nu$. Then, the lowest order relativistic
relaxation time approximation reads
\begin{equation}
\Phi-\Phi_{E}=\nu^{-1}q^{-1}p^{\mu}\partial_{\mu}^{x}\Phi_{E}.
\end{equation} We  obtain $\nu$ if we can calculate the time evolution
(125). We are able to do this only in the high energy limit
$q^{-1}p^{2}\rightarrow 0$. Such a limit is equivalent to the
limit $m\rightarrow 0$ in eq.(76). The time evolution (125) can be
calculated exactly \cite{habapreprint}\cite{habamod}in this limit
and the relaxation time approximation is achieved with
$\nu=\beta\kappa^{2}$. In general, in order to obtain a local in
time fluid equation we need a local in time relativistic
approximation to the solution (125).
 The result of such an approximation can
be inferred from the relativistic invariance. The relativistic
invariant (local in time) approximation to eq.(125) must be of the
form
\begin{equation}\begin{array}{l}
\Gamma(p,x)=1+\partial_{\mu}^{x}w_{\nu}(x)(a_{0}p^{\mu}p^{\nu}+a_{1}w^{\mu}p^{\nu})
+a_{2}p^{\mu}\partial_{\mu}\chi\cr+a_{3}p^{\mu}\partial_{\mu}\beta+a_{4}w^{\mu}\partial_{\mu}\chi
+a_{5}w^{\mu}\partial_{\mu}\beta,\end{array}\end{equation} where
$a_{j}(\tau,pw,p^{2})$ are scalars depending on the available
Lorentz scalars $\tau,pw$ and $p^{2}$. In the relaxation time
approximation we have
\begin{displaymath}
a_{0}=-(\nu pw)^{-1}
\end{displaymath}
and $a_{j}=0$ for $j>0$.

Inserting this $\Gamma$ in eqs.(18) and (105) we can  obtain a
formula for the current, from eq.(107) the formula for the energy
momentum. We could derive the general tensor form of these
expressions. Subsequently, from the  equations of conservation of
the current and the energy momentum we obtain differential
equations for the velocity, temperature and the density. The
calculations have been performed explicitly in the relaxation time
approximation to the Boltzmann equation in the classic paper
\cite{relax}. We would like to concentrate here on the appearance
of the viscosity (which is of interest for high energy physics
\cite{redlich}) in the approximation (125). For this purpose it is
sufficient if we restrict ourselves to $\beta=const$ and
$\chi=const$ . In such a case\begin{equation}\begin{array}{l}
\rho(\Phi)v_{\mu}=
n_{E}\rho_{E}w_{\mu}+c_{1}\rho_{E}w^{\nu}\partial_{\nu}^{x}w_{\mu}(x)+c_{2}\rho_{E}w_{\mu}\partial_{\nu}^{x}w^{\nu}(x)
\end{array}\end{equation} where  $c_{1}$ and $c_{2}$
can be calculated from eqs.(18) and (105) as integrals over $p$
depending on the functions $a_{0}$ and $a_{1}$. In the case of the
relaxation time approximation  the constants $c_{1}$ and $c_{2}$
are determined by the integrals
\begin{equation} s^{\mu_{1}....\mu_{n}}=\int dm^{2}s(m^{2})\int
dp\delta(p^{2}-m^{2})(pw)^{-1}\Phi_{E}p^{\mu_{1}}....p^{\mu_{n}}
\end{equation}calculated in \cite{relax}. The tensor on the rhs of (130) is expressed by $w^{\mu}$ and $\eta^{\mu\nu}$ .
 The energy momentum tensor has the form
\begin{equation}\begin {array}{l}\theta^{\mu\nu}=\theta^{\mu\nu}_{E}+\int
d\sigma p^{\mu}p^{\nu}(\Gamma-1)\Phi_{E}
=(E+\Pi_{E})w^{\mu}(x)w^{\nu}(x)
-\eta^{\mu\nu}\Pi_{E}\cr+\delta_{0}(\partial^{\mu}_{x}
w^{\nu}(x)+\partial^{\nu}_{x}
w^{\mu}(x))+\delta_{4}\eta^{\mu\nu}\partial_{\alpha}^{x}w^{\alpha}
-\delta_{3}w^{\mu}(x)w^{\nu}(x)\partial_{\alpha}^{x}w^{\alpha}(x)
\cr-\delta_{1}(w^{\mu}(x)w^{\alpha}(x)\partial_{\alpha}^{x}w^{\nu}(x)+w^{\nu}(x)
w^{\alpha}(x)\partial_{\alpha}^{x}w^{\mu}(x)),\end{array}\end{equation}
where $E$ and $\Pi_{E}$ are the energy and pressure of the
J\"uttner gas (they are constants if $\chi$ and $\beta$ are
constants in eq.(115)). $\delta_{j}$ can be calculated if the
functions $a_{0}$ and $a_{1}$ are known (in the relaxation time
approximation we determine $\delta_{j}$ from integrals of the form
(130)). We can express $u$ by $w $ from eq.(129)
\begin{equation}\begin{array}{ l} u^{\mu}=w^{\mu}
+c_{1}n_{E}^{-1}w^{\nu}\partial_{\nu}^{x}w^{\mu}.\end{array}
\end{equation}
We can see that till the
 terms first order in gradients we have
\begin{equation}
\partial_{\alpha}^{x}u^{\mu}=\partial^{x}_{\alpha}w^{\mu}
\end{equation}
and
\begin{equation}\begin{array}{l} w^{\mu}=u^{\mu}
-c_{1}n_{E}^{-1}u^{\nu}\partial_{\nu}^{x}u^{\mu}.\end{array}
\end{equation}

Now, we can replace $w $ in  $\theta^{\mu\nu}$ by $u$ and its
derivatives. We have
\begin{equation}\begin{array}{l}
\theta^{\mu\nu}=(E+\Pi_{E})u^{\mu}u^{\nu}-\eta^{\mu\nu}\Pi_{E}+\sigma^{\mu\nu}_{E}\end{array}\end{equation}
where
\begin{equation}\begin{array}{l}\sigma^{\mu\nu}_{E}=
-\sigma_{1}(
u^{\nu}u^{\sigma}\partial_{\sigma}^{x}u^{\mu}+u^{\mu}u^{\sigma}\partial_{\sigma}^{x}u^{\nu})
+\delta_{0}(\partial^{\mu}_{x} u^{\nu}+\partial^{\nu}_{x} u^{\mu})
 -\delta_{3}u^{\mu}u^{\nu}\partial_{\alpha}^{x}u^{\alpha}+
\delta_{4}\eta^{\mu\nu}\partial_{\alpha}^{x}u^{\alpha}.\end{array}\end{equation}
Here
\begin{equation}\sigma_{1}=\delta_{1}+(E+\Pi_{E})c_{1}n_{E}^{-1}
\end{equation}The current conservation and  eq.(133) give (when $\chi$ and
$\beta$ are kept constant in eq.(115))
\begin{equation}\partial_{\mu}^{x}u^{\mu}=0.
\end{equation}
Eq.(114) rewritten in terms of $\sigma^{\mu\nu}_{E}$ (136) with
$w^{\mu}$ (134) expressed by $u^{\mu}$ reads
\begin{equation}\begin{array}{l} \epsilon
u^{\mu}\partial_{\mu}u_{\alpha}=-
h_{\alpha\nu}(u)(\delta_{0}\partial^{\mu}_{x}h_{\mu\sigma}\partial^{\sigma}_{x}
u^{\nu}-(\sigma_{1}-\delta_{0})(u^{\rho}\partial^{x}_{\rho})^{2}u^{\nu}),
\end{array}\end{equation}where
\begin{equation}
\epsilon= E+\Pi_{E}
+\beta\kappa^{2}(-3\Pi_{E}c_{1}n_{E}^{-1}+\delta_{0}-\delta_{1}).
\end{equation}In eq.(139) we have
omitted the squares of the gradients of $u^{\mu}$. This is the
relativistic Navier-Stokes equation with the shear viscosity
$\delta_{0}$. There is  no gradient of the pressure and no bulk
viscosity in eq.(139) because in the state (115) (with constant
$\chi$ and $\beta$ ) in our approximation the fluid is
incompressible and has constant pressure.

 We can see that the friction changes the value of $E
+\Pi_{E}$.  Apart from this minor change we obtain the same
hydrodynamics which we could have derived from the diffusion
theory of sec.3 with $f^{\mu}=0$ (preserving the conservation of
the energy momentum;in the equation for energy balance the
friction would appear).  In our definition of $u^{\mu}$ as
proportional to the current we have chosen the Eckart convention
\cite{eckart}. This means that in general the energy momentum
tensor describes a certain heat flow. We  would need
$\sigma_{1}=\delta_{0}$ to eliminate the heat flow. We do not
expect such an equality to be satisfied for the same reason as in
Weinberg's discussion \cite{wein} of the Thomas model when the
general decomposition of the energy momentum depends on the
definition of the temperature. For a complete discussion of the
heat flow we would need  $x$-dependent $\chi$ and $\beta$ in
eq.(115). If there is no heat flow ( $\sigma_{1}=\delta_{0}$) then
in the fluid rest frame we obtain the Navier-Stokes term
$\triangle u^{j}$ in the fluid equation (139). We cannot formulate
a closed hydrodynamic scheme in the sense of Israel and Stewart
\cite{israel} in the approximation which we are applying in this
section, in order to check its causality and thermodynamic
stability \cite{lindlom}. For this purpose we would have to
discuss the hydrodynamic evolution of the general state (115) to
higher orders of perturbation.

\section{Summary}
The assumption of a continuous mass spectrum enables us to work in
an explicitly covariant way treating all components of the momentum
on the same footing. We have studied the possible relativistic
diffusions either by searching positive definite diffusion matrices
or approximating random dynamics by a diffusion. The latter method
shows that the relativistic diffusion really arises in physical
systems. We have studied in more detail the random electromagnetic
field applied in quantum optics for a description of photons in
cavities at finite temperature. We identified the non-zero
expectation value of the electric current as a possible source of
friction leading to the J\"uttner equilibrium. The resulting
diffusion may have an application to relativistic streams of
particles encountered in astrophysics \cite{jokpi} and in
high-energy physics \cite{rupp}.  In secs.7-9 we have shown how to
apply standard methods (widely used in  the case of the Boltzmann
equation) for a statistical and hydrodynamic description of a stream
of relativistic diffusing particles satisfying the diffusion
equation. In this way we could relate methods based on the diffusion
equation \cite{rupp} to the ones using the hydrodynamic equations in
heavy ion physics \cite{rom}\cite{baier}. The relativistic diffusion
equation can also supply new ways of approaching the problems of
relativistic statistical physics and relativistic hydrodynamics by
means of methods applying differential equations rather than the
integral equations of the Boltzmann type.


\begin{thebibliography}{99}\bibitem{pitaj} E.M. Lifshits and L.P. Pitaevskij,
Physical Kinetics,Pergamon Press,1981
\bibitem{chandra}S. Chandrasekhar, Rev.Mod.Phys.{\bf 15},1(1943)
\bibitem{lopuch}J.
Lopuszanski, Acta Phys.Pol.{\bf 12},87(1953)
\bibitem{hakim}R. Hakim, J.Math.Phys.{\bf 9},1805(1968)\bibitem{deb1}C. Chevalier and F. Debbasch,
 AIP Conf.Proc.{\bf
913},42(2007)
\bibitem{hang}J. Dunkel and P. H\"anggi,Phys.Rep.{\bf 471},1(2009)
\bibitem{kampen}N. Van Kampen, Stochastic Processes in Physics and
Chemistry, 3rd Edition,North Holland,2007
\bibitem{fish}Y. Krivolapov and S. Fishman, arXiv:1203.6895
\bibitem{kesten}H. Kesten
 and G.C. Papanicolaou, Commun.Math.Phys.{\bf 78},19(1980)\bibitem{schay} G.Schay,PhD thesis,Princeton University,1961
\bibitem{dudley} R.Dudley, Arkiv for Matematik,{\bf 6},241(1965)
\bibitem{deb3}M. Rigotti and F.Debbasch, Journ.Math.Phys.{\bf 46},103303(2005)
\bibitem{deb2}C. Chevalier and F. Debbasch, J.Math.Phys.{\bf 49},043303(2008)


\bibitem{mrow}S. Mrowczynski, Ann.Phys.(N.Y.){\bf 169},48(1986)
\bibitem{zachar} P. Carruthers and F. Zachariasen,
Phys.Rev.{\bf D13},950(1976)\bibitem{cmb} P.J.E. Peebles and D.T.
Wilkinson, Phys.Rev.{\bf 174},2168(1968)\bibitem{kujawski}

R.K. Pathria, Proc.Phys.Soc.{\bf 88},791(1966)

 J.H. Eberly and A.
Kujawski, Phys.Rev.{\bf 155},109(1967)
 \bibitem{matsui}
 T. Matsui, B. Svetitsky and L.D. McLerran, Phys.Rev.{\bf D34},783(1986)

\bibitem{juttner}F. J\"uttner, Ann.Phys.(Leipzig){\bf 34},856(1911)
\bibitem{kubo1}R. Kubo, J.Math.Phys.{\bf 4},174(1962)
\bibitem{kubo2}R. Kubo, M. Toda and N. Hashitsume,
 Statistical Physics II. Nonequilibrium Statistical Mechanics, Springer,
Berlin,1985
\bibitem{habapre}Z. Haba, Phys.Rev.{\bf E79},021128(2009)
\bibitem{habaphysica}Z. Haba, Physica {\bf A390},2776(2011)
\bibitem{cerci}C. Cercignani and  G.M. Kremer, The Relativistic
Boltzmann Equation:Theory and Applications, Springer,Berlin,2002
\bibitem{chapman}T.G. Cowling and S. Chapman, The Mathematical
Theory of Non-uniform Gases, Cambridge,1970
\bibitem{huang}K. Huang, Statistical Mechanics, Wiley, New
York,1963
\bibitem{eells}J. Eells and J.H. Sampson, Amer.Journ.Math.{\bf
86},109(1964)
\bibitem{gelfand}I. M. Gelfand and N.Y.Vilenkin, Generalized
Functions, Vol.4,Academic, New York, 1964
\bibitem{operators} M.E. Taylor, Pseudodifferential Operators,
Princeton University Press,1981
\bibitem{calo2} S. Calogero, J.Cosm.Astro.Phys.{\bf 11},016(2011)

\bibitem{habamod}Z.Haba,Mod.Phys.Lett.{\bf A25},2683(2010)

\bibitem{landau}
L.D. Landau and E.M. Lifshits, Field Theory, Pergamon Press,
 New
York, 1981

\bibitem{habajpa}Z.Haba, Journ.Phys.{\bf A44},335202(2011)

\bibitem{calo}J.A. Alcantara and S. Calogero, Kinetic and
Related Models, {\bf 4},401(2011)
\bibitem{eckart}C. Eckart, Phys.Rev.{\bf 58},919(1940)
\bibitem{landaufluid} L.D. Landau and E.M. Lifshits,
Fluid Mechanics,Addison-Wesley,1958
\bibitem{rom}P. Romatschke, arXiv:0902.3663
\bibitem{is}W. Israel, J.Math.Phys.{\bf 4},1163(1963)

\bibitem{groot}S.R. de Groot, W.A. van Leeuwen and Ch.G. van Weert,  Relativistic
Kinetic Theory, North Holland,1980
\bibitem{kremer}G.M. Kremer,Continuum Mech.Thermodyn.{\bf
9},13(1997)\bibitem{habapreprint}Z.Haba, arXiv:0911.3126
\bibitem{relax}J.L. Anderson and H.R. Witting, Physica {\bf
A74},466(1974), {\bf A74},489(1974)
\bibitem{redlich}P. Romatschke and U. Romtschke,
Phys.Rev.Lett.{\bf 99},172301(2007)\bibitem{wein}S. Weinberg,
Astrophys. J.{\bf 168},175(1971)
\bibitem{israel}W.Israel, Ann.Phys.(N.Y.){\bf 100},310(1976)

W. Israel and J.M. Stewart, Ann.Phys.(N.Y.){\bf 118},341(1979)
\bibitem{lindlom}W.A. Hiscock and L.Lindblom, Phys.Rev.{\bf
D31},725(1985),{\bf D35},3723(1987)


\bibitem{jokpi}J.R. Jokpii,Astrphys.J.{\bf 146},480(1966)\bibitem{rupp}R. Rapp and H. Van Hees, in Quark-Gluon
Plasma 4, ed. by R.C. Hwa and Xin-Nian Wang, World Scientific, 2010;
arXiv:0903.10961


\bibitem{baier} R.Baier, P. Romatschke and A.U. Wiedemann,

Phys.Rev.{\bf C73},064903(2006)








\end{thebibliography}
\end{document}